\documentclass[aps,prl,twocolumn,superscriptaddress,groupedaddress]{revtex4}
\usepackage{subfig}
\usepackage{graphicx}  
\usepackage{dcolumn}   
\usepackage{bm}        
\usepackage{amssymb}   
\usepackage{slashed}
\usepackage{graphicx}				
\usepackage{amsmath}
\usepackage{mathtools}
\usepackage{tikz}
\usetikzlibrary{arrows,backgrounds}
\usepackage[all]{xy}
\usepackage{yfonts}
\usepackage{color}

\newcommand{\ket}[1]{\ensuremath{\left|#1\right\rangle}}

\begin{document}
\title{Entanglement, space-time and the Mayer-Vietoris theorem}
\author{Andrei T. Patrascu}
\address{University College London, Department of Physics and Astronomy, London, WC1E 6BT, UK}

\begin{abstract}
Entanglement appears to be a fundamental building block of quantum gravity leading to new principles underlying the nature of quantum space-time. One such principle is the ER-EPR duality. While supported by our present intuition, a proof is far from obvious. In this article I present a first step towards such a proof, originating in what is known to algebraic topologists as the Mayer-Vietoris theorem. 
The main result of this work is the re-interpretation of the various morphisms arising when the Mayer-Vietoris theorem is used to assemble a torus-like topology from more basic subspaces on the torus in terms of quantum information theory resulting in a quantum entangler gate (Hadamard and c-NOT).
\end{abstract}

\maketitle
\section{1. Introduction}
The origin of entanglement lies within basic quantum mechanics [1]. However, there is no doubt today that there is a connection between quantum entanglement and the emergence of space-time [2], [3]. At a very intuitive level the statement behind the newly discovered ER-EPR duality [4] is very appealing. The connection between space-time topology and entanglement however remains an unproved conjecture. The ideas behind it have already been mentioned in [5], [6], [10] and several conclusions have been extracted in [7], [8]. The new formulation of the ER-EPR duality basically reminds us that the statistical correlation between space-like separated regions associated to generic quantum field theories may have a topological interpretation as well. However, the algebraic topological implications of the ER-EPR statement have only marginally been explored [9], [11]. In this article I will connect quantum entanglement to space-times with non-trivial topology by means of the Mayer-Vietoris sequence [12]. The main tool used will be quantum field theory. This is not the most natural tool for describing quantum information problems. However, standard quantum information problems are usually not analyzed in curved or topologically non-trivial space-time. Focusing exclusively on a basic quantum mechanical approach as is done in standard quantum information theory may prove to be unsustainable when space-time horizons and non-trivial space-time topologies arise. Therefore a brief introduction in the algebraic properties of generic quantum field theories will be presented. On the quantum information side, the standard definition of a qubit will have to be extended in order to be meaningful in quantum field theory. The approximate way in which such a quantum information entity may have sense in the context of quantum field theory will be briefly described. The other important component of this paper, namely entanglement, must also be introduced in the proper context of quantum field theories. This has been done before by means of entanglement entropy. This concept had an important impact on various branches of physics. For example some phases of matter need to be characterized by their pattern of entanglement rather than the conventional order parameters [20], [21]. Quantum entanglement has already been used to characterize various properties of quantum field theory. For example some questions related to the nature of the renormalization (semi)-group have been answered in this way in [22]. The association of entanglement entropy to the geometric structure of the bulk space in the AdS/CFT duality has been the subject of research like [23]. The generalization of these ideas has led to the calculation of the entanglement entropy of a conformal field theory for a subsystem with an arbitrary boundary [24]. The next generalization, involving global aspects of the bulk space has been discussed in [25] where the area encoding the entanglement entropy which entered the bulk space was considered to encircle a non-trivial cycle of the bulk topological space (e.g. a great circle of a torus). In this article the topological properties of the space will play a fundamental role as well, although the entanglement will be described by quantum field theoretical generalizations of measures like the Bell inequalities. In this work I will focus mainly on bipartite entanglement leaving the multipartite case for a future research. 
\par The structure of this article is as follows. In the second chapter I will introduce the basics of relativistic algebraic quantum field theory in flat and curved space-time, focusing on the definition of entanglement and qubits in this context. An intuitive justification for the use of (co)homology groups for the classification of curved space-time quantum field theoretical qubits will also be given.
In the third chapter I will provide a link between the various sets of observables, the topology of space-time and the presence of entanglement. I will also provide details about the geometry of an ER-bridge in terms of Kruskal coordinates as well as the main geometrical and topological context of this article.
In the fourth chapter a pedagogical overview of Mayer-Vietoris theorem and the way of thinking implied by it will be presented. 
In the fifth chapter I present the main results of this article in the form of two theorems and a corollary.
In the sixth chapter I will connect the new insights offered by the Mayer-Vietoris theorem to the concept of quantum entanglement for flat, topologically trivial space-time and for a space-time connected via an ER-bridge. I also show how entanglement is a natural result of the application of the Mayer-Vietoris theorem. By means of basic quantum information techniques it will be seen that the maps involved in the Mayer-Vietoris theorem are analogous to the entangler gate (Hadamard followed by c-NOT). Reversely, I will show that disconnected patches of space-time with entanglement between them can be reformulated as regions connected by means of ER bridges when certain non-trivial coefficient systems in (co)homology are being used. Moreover, it is worth noting that universal coefficient theorems connecting ordinary and generalised cohomologies play the role of maps relating theories based on point-like structures and theories based on extended structures. The extended structures play the role of natural regularisers in the same way as strings do. This aspect of generalised cohomology theory is only briefly mentioned in this article in order to reassure the reader worried that potential divergences from quantum field theories could ruin the discussion. The detailed discussion of the connection between renormalisation of operator product expansions and generalised cohomology is left for an upcoming set of articles.
In the seventh chapter, I present a generalisation by means of the Reeh-Schlieder theorem. I connect the maps arising in the Mayer-Vietoris sequence to the state-operator correspondence in generic quantum field theory and briefly show how this would particularize for conformal field theories by means of homologies with twisted coefficients.
Finally, I will provide some conclusions as well as new directions of research.

\section{2. Relativistic algebraic quantum field theory}
In order for this article to be self-contained, a discussion about the meaning of entanglement in quantum field theory is required.
Indeed, like in basic quantum mechanics, a relatively good indicator for entanglement is the violation of Bell's inequalities. This must however be formulated in the context of generic quantum field theories. Two mathematically rigorous formulations exist: one based on quantum fields satisfying the Wightman axioms and the other one based on local algebras satisfying the Haag-Kastler-Araki axioms. Both allow consistent descriptions of entanglement. 

In the local algebraic description of quantum field theory, Bell's inequalities concern results of correlation experiments involving measurements on two subsystems. Such experiments can be characterized according to [13] by the so-called correlation dualities. These represent a set of three objects, $(\hat{p},\mathcal{A},\mathcal{B})$. $\mathcal{A}$ and $\mathcal{B}$ being real vector spaces with a specific vector ordering defined on them and having a well defined identity $id=1$. $\hat{p}$ is a bilinear function $\hat{p}:\mathcal{A}\times\mathcal{B}\rightarrow \mathbb{R}$. The observables of one such subsystem are represented by partitions of the identity in the respective subsystem i.e. $\{a_{i}| i\in I\}$, $\sum_{i}a_{i}=1$, $a_{i}\geq 0$, $\forall i\in I$. Every $i\in I$ is interpreted as a possible outcome of the measurement of an observable $a_{i}$. The probability of the joint occurrence of two outcomes $i\in I$ and $j\in J$ in the respective two subsystems will then be by definition $\hat{p}(a_{i},b_{j})$. Using this definition the Bell correlation is defined as 
\begin{widetext}
\begin{equation}
\beta(\hat{p},\mathcal{A},\mathcal{B})=\frac{1}{2}sup(\hat{p}(x_{1},y_{1})+\hat{p}(x_{1},y_{2})+\hat{p}(x_{2},y_{1})-\hat{p}(x_{2},y_{2}))
\end{equation}
\end{widetext}
the supremum being taken over all $x_{i}\in\mathcal{A}$ and $y_{i}\in\mathcal{B}$. The expression for the Bell equality is then $\beta(\hat{p},\mathcal{A},\mathcal{B})=1$ which we expect to be violated. When the vector spaces $\mathcal{A}$ and $\mathcal{B}$ modeling the observables of the considered subsystems are in fact $C^{*}$ algebras (like in quantum mechanics) the Bell correlation satisfies the inequality $\beta(\hat{p},\mathcal{A},\mathcal{B})\leq\sqrt{2}$. 

When dealing with relativistic quantum field theory the basic structure is an assignment to each open region $\mathcal{O}\in \mathbb{R}^{4}$ of a $C^{*}$-algebra $\mathcal{A}(\mathcal{O})$ of norm-closed bounded operators on some Hilbert space. This assignment must satisfy certain axioms originating in physics. 

First if there are two regions $\mathcal{O}_{1}\subseteq \mathcal{O}_{2}$ then the associated algebras also satisfy $\mathcal{A}(\mathcal{O}_{1})\subseteq \mathcal{A}(\mathcal{O}_{2})$. Therefore, each $\mathcal{A}(\mathcal{O})$ is a subalgebra of the $C^{*}$-algebra $\mathcal{A}$ generated by 
$\bigcup\limits_{\mathcal{O}\subset \mathbb{R}^{4}}\mathcal{A}(\mathcal{O})$. 

Second, in order to define the flat relativistic space-time, Poincare covariance must be obeyed. Therefore, for flat space-times there must exist a representation $\{\alpha_{\lambda}|\lambda\in \mathcal{P}^{\dagger}_{+}\}$ of the identity connected component $\mathcal{P}^{\dagger}_{+}$ of the Poincare group by a group of automorphisms on $\mathcal{A}$ such that $\alpha_{\lambda}(\mathcal{A}(\mathcal{O}))=\mathcal{A}(\mathcal{O}_{\lambda})$ where $\mathcal{O}_{\lambda}$ is the image of $\mathcal{O}$ under the transformation corresponding to $\lambda$. 

Third, if $\mathcal{O}_{1}$ is spacelike separated from $\mathcal{O}_{2}$ then every element of the algebra $\mathcal{A}(\mathcal{O}_{1})$ commutes with every element of the algebra $\mathcal{A}(\mathcal{O}_{2})$. This assures the existence of a notion of locality. 
It is important to make a clear distinction between what I call locality in this article, namely the property that observables in spacelike separated regions commute, and another, weaker definition of locality used sometimes in quantum information theory, focusing mostly on the quantum fields or their simpler analogues, the wavefunctions. Indeed, apparent non-local effects resulting from wavefunction superpositions or quantum field correlations are not truly non-local according to the definition of this article.  

Finally, there exists a physical, faithful i.e. one-to-one representation $\pi$ of $\mathcal{A}$ on a separable Hilbert space $\mathcal{H}$ such that on $\mathcal{H}$ there is a nontrivial strongly continuous unitary representation $U(\mathcal{P}^{\dagger}_{+})$ of the universal covering group of the Poincare group $\mathcal{P}^{\dagger}_{+}$ satisfying first, $U(\lambda)\pi(A)U(\lambda)^{-1}=\pi(\alpha_{\lambda}(A))$ for each $A\in \mathcal{A}$, $\lambda\in \mathcal{P}^{\dagger}_{+}$, and second, the generators $\{P_{\mu}\}_{\mu=0}^{3}$ of the translation subgroup satisfy the condition $P_{0}^{2}-P_{1}^{2}-P_{2}^{2}-P_{3}^{2}\geq 0$ and $P_{0}\geq 0$ where $P_{0}$ is the generator of time translations. Self adjoint elements $A\in \mathcal{A}(\mathcal{O})$ of the local algebras are interpreted as observables which are measurable in the corresponding space-time region $\mathcal{O}\subset \mathbb{R}^{4}$. 
A positive, normalized linear functional $\phi$ on the $C^{*}$-algebra $\mathcal{A}$ is supposed to correspond to a physical state of the system whose local observables are represented by the net $\{\mathcal{A}(\mathcal{O})\}$. For such a state $\phi$ and an observable $A\in \mathcal{A}(\mathcal{O})$, $\phi(A)$ is considered to be the expected value of the observable $A$ of the statistical system that has been prepared in the state $\phi$. 

If $\mathcal{A}$ and $\mathcal{B}$ are commuting $C^{*}$-algebras and $\phi$ is a state on a $C^{*}$-algebra $\mathcal{C}$ containing both $\mathcal{A}$ and $\mathcal{B}$ then $(\phi, \mathcal{A}, \mathcal{B})$ determines a correlation duality $\hat{p}(A,B)=\phi(AB)$ for each $A\in\mathcal{A}$ and $B\in\mathcal{B}$. Therefore if $\phi$ is a state on an algebra $\mathcal{A}$ generated by a net of local algebras $\{\mathcal{A}(\mathcal{O})\}$ and if $\mathcal{O}_{1}$ and $\mathcal{O}_{2}$ are any two spacelike separated regions in space-time then $(\phi, \mathcal{A}(\mathcal{O}_{1}), \mathcal{A}(\mathcal{O}_{2}))$ is a correlation duality.

In the alternative formulation based on the Wightman axioms we employ so called quantum fields i.e. operator valued distributions $\phi$ on space-time which act on the physical state space. These fields are then integrated with test functions $f$ having support in a given region $\mathcal{O}$ of space-time $\phi[f]=\int d^{4}xf(x)\phi(x)$. The resulting objects form under the operations of addition, multiplication and hermitian conjugation a polynomial *-algebra $\mathcal{P}(\mathcal{O})$ of unbounded operators. 


Both approaches however assume Poincare invariance and therefore must be replaced with local Lorentz invariant formulations when space-time is curved. Moreover, if we want to connect quantum field theory to quantum information theory, we need a sufficiently accurate description of a qubit. Given a Hilbert space, a qubit can be physically realized as any two dimensional subspace of that Hilbert space. Such realizations however will often not be localized in space. We can restrict ourselves to approximately well localized realizations and represent the qubit as a two dimensional quantum state attached to a single point in space. If we want to ensure relativistic invariance we notice that there are no finite dimensional faithful unitary representations of the Lorentz group. For flat space-time we can go to the Wigner representations. These provide us with unitary and faithful but still infinite dimensional representations of the Lorentz group. These representations strongly rely on the symmetries of Minkowski space and in particular on the inhomogeneous Poincare group. The basis states are taken to be eigenstates of the four-momentum operator such that $\hat{P}^{\mu}\ket{p,\sigma}=p^{\mu}\ket{p,\sigma}$ where $\sigma$ refers to some discrete degree of freedom i.e. a spin or a polarization. To obtain a physical two-dimensional quantum state we may restrict ourselves to a specific momentum eigenstate $\ket{p,\sigma}$ of fixed $p$. The remaining degrees of freedom will then be discrete. However, when we go from flat to curved space-time we loose the translational symmetry and therefore the momentum eigenstates $\ket{p,\sigma}$. We still have local Lorentz invariance. A qubit must still be understood as a two-level quantum system with the property of being spatially well localized. The history of such a localized quantum system is a sequence of two dimensional quantum states $\ket{\psi(\lambda)}$ each associated to a point $x^{\mu}(\lambda)$ on the worldline parametrized by $\lambda$. Each quantum state in this sequence $\ket{\psi(\lambda)}$ must be thought as belonging to a distinct Hilbert space $\mathcal{H}_{x(\lambda)}$ attached to each point $x^{\mu}(\lambda)$ of the trajectory. The parallel transport is then a sequence of infinitesimal Lorentz transformations acting on the quantum state and this sequence is in general path dependent. Therefore, in general it is not possible to compare quantum states associated with distinct points in space-time. As a consequence it is not meaningful to say that two quantum states associated to distinct points in space-time are the same. We may however use quantum teleportation and entangled states to define what means "the same" in the context of curved space-time. Therefore the whole sequence of quantum states attached to points along a worldline describing the history of $\ket{\psi(\lambda)}$ will be called a quantum field theoretical qubit.

Similar methods have already been employed in [54] and [55] with the aim of generalising the discussions focused on quantum information theory to quantum field theory and high energy physics in curved spacetime. It is well known that the applicability domain of the above mentioned constructions is limited by the quantum field theoretical structure of our underlying theory. This has already been noticed in [54]. Such a formalism implies the spatial localisation of qubits, a property that cannot be exactly defined in quantum field theory. Therefore the situation in which spacetime curvature is comparable with the wavepacket width of the qubits will become problematic. These situations may be avoided by assuming that extreme curvature scales do not occur in our problem. It is also well known that quantum field theory in curved spacetime does not have a unique vacuum state and therefore the particle number is susceptible to ambiguities. In fact such ambiguities result in particle creation around black holes. Particle creation in the form of entangled pairs will result in an increase of the topological complexity of the problem. This will of course result in yet another limitation to a strictly topological description in terms of ordinary cohomology theory. Such a simple minded cohomology theory is based on the acceptance of the dimension axiom of the Eilenberg-Steenrod axiom set defining ordinary cohomology theories. This axiom implies that the (co)homology of a point is non-zero only in degree zero and there, it is isomorphic to $\mathbb{Z}$ i.e. the point is a simply connected object of zero dimension (for a detailed definition see [57]). This fact is challenged first in a weak sense in quantum field theory where exact localisation is redefined in a distributional sense and next, in string theory, where we add the structure of extended objects to the mathematically point-like particles. Therefore by using cohomology with non-trivial coefficients and the universal coefficient theorem it is possible to move back and forth between point-like representations and extended representations avoiding, therefore, most of the divergences of ordinary quantum field theories.  As will be seen later in this article, if one considers the topology of spacetime itself not as an {\it a-priori} given fact, but instead as having a visibility which depends on the choice of the coefficient structure in cohomology [17], the situation will become less paradoxical, albeit it may have to be described by means of a special type of linear topological algebraic tools. There is a certain inclination towards calling such tools "non-linear". This would be incorrect as the normal linear structure we know from quantum mechanics is preserved. There is no local linear observable that can detect the topology of spacetime with absolute certainty as there is no local linear observable to be associated to entanglement. This has been associated to the fact that it is not possible to have projector operators that would project onto a subspace unless that subspace is closed under superposition [56]. An attempt to project onto the set of all entangled states will fail due to the fact that the set of all entangled states is not closed under linear superposition. If such a projector would be formed it would inevitably project onto the entire Hilbert space of all states [56]. When more than a single isolated observer is employed (i.e. observers are allowed to exchange signals), reference [56]  shows that there only exists the possibility to determine if the region behind the horizon contains certain particular wormhole states i.e. the methods can sometimes reveal the existence of a wormhole, but cannot rule out its presence definitively. The visibility of topology strongly depends on the coefficient structure in cohomology. I noticed this in [17] and it can further be interpreted in the sense that as long as the probing of topology is done via individual, localised (point-like) objects, the topology of a space cannot be completely revealed. When generalised cohomology with non-trivial coefficients is employed, there is a tradeoff between the properties of topology that can and can not be detected [17]. I call the coefficient structure in cohomology a "theoretical measurement tool" [57] capable of revealing some topological properties while hiding others. The coefficient structure in cohomology is seen as an extension from the ordinary cohomology in the sense of adding structure to the mathematical point. Such structure can be interpreted as an extended object or as multiple observers that may give indications about the possibility of non-trivial topology. When the coefficients are chosen such that they partially reveal the topology of spacetime (or otherwise stated, the entanglement) they are not associated to local linear observables. Linearity however can be preserved provided one takes into account multiple observers or extended probing structures. This amounts to saying that while observers can check for the presence or absence of specific ER bridge configurations, there is no projection operator (observable) onto the entire family of wormhole geometries, just as there is no projection operator onto the family of entangled states [56]. While linearity is preserved, the "probing device" gains additional structure given by the coefficient structure, that may reveal more details about the topology. The main advantage of combining non-trivial coefficient structures with linear tools provided by homological algebra is that (at least formally) problems related to divergences of quantum field theories can be avoided. Interpreting operator product expansions and the therein involved renormalisation in terms of generalised cohomology theory is a research project on its own and will be the subject of a set of future articles. At this point however we can continue with the discussion assuming that universal coefficient theorems relating zero dimensional point-like cohomology theories to extended object representations play the same role string theory plays in offering a natural regularisation of the theories. Generalisations to multipartite entanglement will be the subject of a future article. A similar situation may occur in the case of accelerated observers. A description of a topological interpretation of the Unruh effect is again left for a future article with a less ordinary algebraic homological focus. Finally, if we wish to describe extreme energy domains, we will have to consider using cohomology with more advanced coefficient structure (potentially elliptic curves and elliptic groups), therefore adding structure to the mathematical points of our space. A particular situation of this type is of course known as the relativistic quantum string, which may be used to probe our ER topology and furthermore, to enlarge it by inducing additional topological structure. Such string-like structure added to our mathematical points also acts as a natural regulariser which allows us to avoid singularities occurring in quantum field theories.


One can of course take a localized qubit in a superposed state and split it up into a spatial superposition transported simultaneously along two or more distinct worldlines and make it recombine at some future space-time region to produce quantum interference phenomena [38]. Such spatial superpositions will still be considered to be localized if the components of the superposition (the two elements of the expectation catalogue) are each well localized around space-time trajectories [39], [40], [41].
Moreover, any qubit can be written as a superposition of states by means of the Hadamard matrix. Therefore any qubit can be written in terms of topological cycles. The classification of such cycles is then naturally based on a (co)homology theory. 

Taking into account the topology of the space, various qubits can be classified according to the possible deformations such worldline cycles may support. For a simply connected space the situation is straightforward. Any such cycle can be continuously deformed to a single worldline without leaving the space. For a $p$-connected space-time with $p\geq 2$ there exist certain classes of worldlines cycles that cannot be continuously mapped into simple worldlines i.e. cannot be rotated back by simple one-qubit Hadamard matrices. Such classes depend on the connectivity of the space and are precisely defined by (co)homology groups. A cycle can also be constructed by taking the tensor product of two qubits. In particular two-qubit states may correspond to two worldline segments which may be connected in various ways. If the two worldlines combined belong to a non-trivial (co)homology group then there exists an obstruction in expressing them independently on the given space-time topology and therefore they may not be considered as separable. At this point this new and original connection between space-time topology and superposed quantum states starts being clear. The rest of this article will go further and connect entanglement to topology by a similar way of thinking. 

\section{3. Relativity of entanglement}
It is not new [14], [15] that the partition of a quantum system into subsystems is dictated by the set of operationally accessible measurements. Given a Hilbert space $\mathcal{H}$ it is possible to either look at it as a bipartite space i.e. $\mathcal{H}_{1}\otimes \mathcal{H}_{2}$ or as an irreducible space $\mathcal{H}$. If the space can be seen as a bipartite space then a tensor product structure exists and this may support entangled states i.e. states that cannot be represented as a direct product of separate states on each of the partitions of the Hilbert space. But what induces the partitioning of a given Hilbert space? It has been argued by [16] that this partitioning is due to the experimentally accessible observables. Therefore an entangled state is only defined as such when the particular experimental setup capable of detecting the associated properties is specified.

However, I reiterated in [17] that observables and quantum states, when described in terms of (co)homology groups (see for example [31], [32], [33] but also the discussion of the previous chapter), are dependent on the coefficient groups used. Indeed, given certain choices of coefficients in (co)homology, observables can merge together becoming undistinguishable. 

Another important aspect is that the use of certain coefficient groups may mask the topological properties of an underlying space. Therefore, topology can be perceived by quantum states and observables only with an accuracy given by the particular coefficient groups in (co)homology. In order to be more specific, take the torus $T_{2}$. Its integral cohomology in dimension $1$ is $H^{1}(T_{2};\mathbb{Z})=\mathbb{Z}\oplus \mathbb{Z}$ and the $0$-dimensional and $2$-dimensional homology groups are each isomorphic to $\mathbb{Z}$. However, the first cohomology group $H^{1}(T_{2};\mathbb{G})$ with coefficients in a group $\mathbb{G}$ is isomorphic to the group of homomorphisms from $\mathbb{Z}\oplus\mathbb{Z}$ to the group $\mathbb{G}$. This group $Hom(\mathbb{Z}\oplus\mathbb{Z},\mathbb{G})$ is trivial if $\mathbb{G}$ is a torsion group. If not, it is a direct sum of copies of $\mathbb{G}\oplus\mathbb{G}$. Hence the torsion of the coefficient group in cohomology determines the visibility of a torus as such. The supplemental information accessible with one coefficient group remains only encoded in the extension $Ext$ that appears in universal coefficient theorems used when changing the coefficient groups. 

Therefore, from the perspective of (co)homology with coefficients and implicitly of quantum states or quantum observables, there exists a duality between toruses and spheres, the relation between the two shapes being given by a particular choice of coefficients. 

It is therefore pertinent to ask what will happen with the entanglement when coefficient groups in cohomology are being chosen such that the space appears to be a torus i.e. when an ER bridge emerges.

At this point it is important to understand what an ER bridge is and how it can be described from a topological point of view. For this I will briefly follow the classical paper by Kruskal [19] and review the concept of maximal extension of the Schwarzschild metric. 
If we start from the well-known Schwarzschild expression for the metric around a center of mass $m(g)$ and use the notation $m^{*}=Gm/c^{2}$ we have 
\begin{equation}
ds^{2}=-(1-2m^{*}/r)dT^{2}+(1-2m^{*}/r)^{-1}dr^{2}+r^{2}d\omega^{2}
\end{equation}
with 
\begin{equation}
d\omega^{2}=d\theta^{2}+sin^{2}(\theta)d\phi^{2}
\end{equation}
With these we observe two types of singularities at $r=0$ and $r=2m^{*}$. While the singularity at $r=0$ is real, the singularity at $r=2m^{*}$ is not. This can be shown by introducing the so called Kruskal coordinates which have the property of being well defined in all regions except the physical singularity. To find this set of coordinates one may seek a spherically symmetric coordinate system in which radial light rays everywhere have the same slope $\frac{dx^{1}}{dx^{0}}=\pm1$ i.e. 
\begin{equation}
ds^{2}=f^{2}(-dv^{2}+du^{2})+r^{2}d\omega^{2}
\end{equation}
By identifying the two metrics defined above and requiring $f$ to depend on $r$ alone and to remain finite and nonzero for $u=v=0$ one finds a set of unique equations of transformation between the exterior of the ``spherical singularity" $r>2m^{*}$ and the quadrant $u>|v|$ in the plane of the new variables [19]
\begin{equation}
\begin{array}{cc}
u=[(\frac{r}{2m^{*}}-1)]^{\frac{1}{2}}exp(\frac{r}{4m^{*}})cosh(\frac{T}{4m^{*}})\\
\\
v=[(\frac{r}{2m^{*}}-1)]^{\frac{1}{2}}exp(\frac{r}{4m^{*}})sinh(\frac{T}{4m^{*}})\\
\\
f^{2}=(32m^{*3}/r)exp(-r/2m^{*})\\
\end{array}
\end{equation}
The new coordinates give an analytic extension $\mathcal{E}$ of that limited region of space-time $\mathcal{L}$ which is described without singularity by Schwarzschild coordinates with $r>2m^{*}$. The metric in the extended region joins smoothly to the metric at the boundary of $\mathcal{L}$ at $r=2m^{*}$. The extended space $\mathcal{E}$ is the maximum singularity free extension of $\mathcal{L}$ that is possible. Every geodesic followed in whichever direction either runs into the barrier of intrinsic singularities at $r=0$ $(v^{2}-u^{2}=1)$ or is continuable infinitely with respect to its natural length. The maximal extension $\mathcal{E}$ has a non-euclidean topology and particularly is one of the topologies considered by Einstein and Rosen (the ER topology). 

Consider therefore a space-time and let it contain a compact region $\Omega$ with non-trivial topology (i.e. the topology of an $n$-torus, a Klein bottle, etc.). As a simplification, the asymptotic regions may be compactified such that the whole picture appears to be isomorphic to a torus. I will consider the compactified and non-compactified objects similarly and I will not start any speculations about the topology of the outer regions (i.e. the large scale topology of the universe) here. For all practical purposes of this article, the ER-bridge will look like $\Omega \cong \mathbb{R}\times \Sigma$ where $\Sigma$ is a 3-manifold with non-trivial topology (i.e. torus, Klein bottle, etc.). When looking at the hypersurfaces $\Sigma$ we have to see them as spacelike in this context. As a slight simplification I will discuss the case of a two-dimensional torus embedded in a three dimensional space in this article. This doesn't affect the generality of the discussion. Going to higher dimensions and to spaces with higher genera will be the subject of a future article where multipartite entanglement will be the main focus. Here, the main subject will be recovering bipartite entanglement from topological considerations alone and therefore the torus $T_{2}$ is sufficient. 

I shall call an ER-bridge as being topologically a torus. The important feature that leads me to this name is that the space-time in this case contains a worldtube (the time evolution of a closed surface) that cannot be continuously deformed into a world line (the time evolution of a point). This is the homotopical definition of a torus. This deformation can however be done on a sphere, and it generates the homotopical definition of a sphere which is equivalent to that of a plane i.e. on both, any closed curve can be homotopically deformed into a point. This is the context in which I will use the terms ``torus" and ``sphere" in this article. 

I showed in the previous chapter that the quantum field theoretical analogue of qubits in curved space-times are to be associated to worldlines. When the qubit is in a superposed state such a worldline can be seen as a cycle. Let me therefore call $\ket{\Psi}$ a qubit associated to the geodesic relating the exterior of the black hole to its interior, which avoids the intrinsic singularity and is continuable indefinitely with respect to its natural length. This would be a qubit state in the context of an ER bridge. It doesn't take too much effort to notice that such a worldline (qubit) is not continuously deformable into a worldline which never enters the horizon in the first place. Also, a superposed qubit which splits between the interior and the exterior of the ER-bridge forms a cycle which cannot be reduced to a point i.e. a cycle which is not a boundary. Also, connecting two non-superposed worldlines we may obtain a worldline around a large cycle of the above defined torus. Such a cycle will also belong to a non-trivial (co)homology. The worldline segments remaining only inside the wormhole or only outside will form elements in a trivial (co)homology. Therefore, qubits, seen as worldlines are classified in terms of (co)homology groups and pairs of qubits may belong to non-trivial (co)homology groups. With this, the connection between quantum information, qubits and homological algebra is established.

\section{4. The Mayer-Vietoris sequence}
One main result connecting algebraic topology and homological algebra is the so called Mayer-Vietoris sequence. Its main underlying idea is that the (co)homology of a given space may be obtained via the (co)homology of some subspaces defined on that space together with the intersection of those subspaces. Otherwise stated, the following sequence is exact 
\begin{widetext}
\begin{equation}
\begin{array}{c}
...\rightarrow H_{n+1}(X)\xrightarrow{\partial_{*}}H_{n}(A\cap B)\xrightarrow{(i_{*},j_{*})}H_{n}(A)\oplus H_{n}(B)\xrightarrow{k_{*}-l_{*}}H_{n}(X)\xrightarrow{\partial_{*}}H_{n-1}(A\cap B)\rightarrow \\
\\
 ... \rightarrow H_{0}(A)\oplus H_{0}(B)\xrightarrow{k_{*}-l_{*}}H_{0}(X)\rightarrow 0\\
 \end{array}
\end{equation}
\end{widetext}
Here $H_{n+1}(X)$ is the homology of the original space $X$, $A$ and $B$ are the subspaces of $X$ chosen to describe the topological properties of the whole space $X$, $H_{n}(A\cap B)$ is the $n$-th homology of the intersection of the two considered subspaces and finally $H_{n}(A)\oplus H_{n}(B)$ is the direct sum of the $n$-th homologies of the considered subspaces. The associated maps are defined as follows:
 the map $i$ includes $A\cap B$ into $A$, $i:A\cap B\hookrightarrow A$, the map $j$ includes $A\cap B$ into $B$, $j:A\cap B\hookrightarrow B$, the map $k$ includes $A$ into $X$, $k:A\hookrightarrow X$ and the map $l$ includes $B$ into X, $l:B\hookrightarrow X$. The map $\partial_{*}$ is a boundary map lowering the dimension of the given group. 
 The symbol $\oplus$ denotes the direct sum of the respective homology groups or modules. 

\par This is a purely mathematical result. However, its implications for physics and most importantly for the construction of a quantum theory of space-time (and implicitly gravity) cannot be ignored. The main statement of Mayer-Vietoris is that the (co)homology of a space with a more complicated topology can be calculated by dividing that space into pieces of known (co)homology and assembling them together in a controlled way. 
The main goal of this article is to show that the formation of a space-time torus induces entanglement via the various maps appearing in the Mayer-Vietoris sequence. Reciprocally, entanglement of two qubits induces a superposition which results in a $p$-connected space-time when the coefficient structures of the associated (co)homology groups are modified accordingly.

\section{5. Mayer-Vietoris and ER-EPR duality}
The main idea behind the ER-EPR duality is that a non-trivial space-time topology can be associated to the entanglement of two patches of space-time in a trivial topology. 
I will not insist on the particular geometry of the space-time in this article as the main idea behind ER-EPR is about topology. As a basic example one can consider a situation in which a black hole forms in a certain region of space-time and it is continued via a hyper-cylinder to another region of space-time where another black hole forms. The process that leads to the formation of such a structure alters the topology significantly. In fact, one may start with a topologically trivial space-time and end up with a topologically non-trivial one. The final configuration in the present context is conventionally called an Einstein-Rosen bridge (short ER bridge). Obviously, no actual information transfer is possible as the wormholes are non-traversable. 

This space can be described as a simple tensorial product of circles, similar to any generic torus. Concretely the space can be written as $T^{n}=S^{1}\times ... \times S^{1}$ i.e. the $n$-fold product of a circle. Quantum states however, as I have shown previously are to be searched in the (co)homology of a given space. 

In order to compute the cohomology associated to quantum states on a non-trivial topology one needs theorems similar to Mayer-Vietoris.

The first step in the construction is to find an open cover of $S^{1}$ (one of the constituent circles of the torus) by two (hyper)-intervals $I_{1}$ and $I_{2}$ such that the intersection $I_{1}\cap I_{2}$ is equal to the disjoint union $J_{1}\bigsqcup J_{2}$ of two smaller intervals. Now, by employing the Mayer-Vietoris sequence for the open cover $U=I_{1}\times T^{n-1}$, $V=I_{2}\times T^{n-1}$ and $U\cap V = (J_{1}\bigsqcup J_{2})\times T^{n-1}$. This leads by induction (assuming integer coefficients) to the homology of the torus 
\begin{equation}
H_{k}(T^{n})=\mathbb{Z}^{(^{n}_{k})}
\end{equation}
where $(^{n}_{k})$ is the binomial coefficient of n choose k. 
What is important to notice in this otherwise standard calculation is the physical interpretation: when the space-time deforms itself so strongly that the topology changes, in order to calculate the associated homology and hence the associated quantum states, we may have to split the space in pieces with easily computable (co)homologies. These are to be associated with unentangled systems in standard quantum mechanics. However, these are never sufficient to compute the actual cohomology. Therefore looking for example only at the two black holes we always miss important topological information. This information is retrieved if we correctly make use of the Mayer-Vietoris theorem and therefore include the (co)homology of the intersection of the homological extensions of the two open covers used in the first place. This intersection may have non-trivial topology and represents the entanglement when looked upon from a quantum mechanical perspective. 
Therefore I now arrive at the main theorems of this work 
\\
\par \textbf{Theorem 1.}
The inclusion map relating the homology of the intersection of two subspaces of the full topological space $X$ to the direct sum of the homologies of the same two subspaces induces a Hadamard-matrix operation which affects the qubit associated to the branch it acts upon. The map which includes the direct sum above into the full topological space $X$ is a c-NOT operation on the branches associated to the two qubits. The global effect of these two maps arising in the Mayer-Vietoris sequence for a torus is the entanglement of the qubits described by the worldlines on the two branches of $X$. 
\\
\par \textbf{Theorem 2.}
Two entangled qubits correspond each to worldlines which, combined, induce the (co)homology of a not simply connected space. Superpositions of the qubits are equivalent to combinations of (co)homology groups as presented via the Mayer-Vietoris theorem for the torus. The apparently disconnected components can be considered (not simply) connected if the coefficient structures in the (co)homologies associated to the respective qubits becomes torsional or cyclical. 
\\
\par \textbf{Corollary 3.}
In general the ER-EPR conjecture is true, the entanglement being in all situations induced by the inclusion maps appearing in the Mayer-Vietoris sequences.
$\flat$
\\

\section{6. Entanglement, inclusion maps and coefficients in (co)homology}

\par In what follows I give proofs of the first theorems and partial evidence for the final corollary. First I will revise some known facts about the entanglement of vacuum and some basic entanglement measures. This will prepare the stage for the discussion in terms of homological deformations of the covering domains and the connectivity of the space-time itself. Finally, the maps of the Mayer-Vietoris sequence required for the construction of a torus will be interpreted in terms of quantum information gates.

\subsection{Bell equality violation and vacuum entanglement measures}
As seen in the chapter referring to relativistic algebraic quantum field theory, a good measure for entanglement is the Bell inequality. More explicitly, states which violate Bell's inequalities are necessarily entangled although states which are entangled may not violate Bell's inequality. Given a quantum system, we may define a pair of algebras, say $(\mathcal{M},\mathcal{N})$, associated to the observables of two subsystems defined each over the space-time regions $\mathcal{O}_{1}$ and $\mathcal{O}_{2}$. A physical state may be defined as $\phi:\mathcal{A}\rightarrow \mathbb{C}$ where $\mathcal{A}$ is an observable algebra with observables defined over a space-time region $\mathcal{O}$. A given such state is called a product state across the pair of algebras $(\mathcal{M},\mathcal{N})$ if $\phi(MN)=\phi(M)\phi(N)$ for all $M\in\mathcal{M}$ and $N\in\mathcal{N}$. In such states the observables of the two subsystems are not correlated and the subsystems are in a sense independent. In terms of quantum field theoretical qubits this translates into the states $\phi(M)$ and $\phi(N)$ belonging to a trivial (co)homology group i.e. there is no obstruction against merging the actions of $M$ and $N$ in the same quantum state. In terms of worldlines, there is a smooth deformation taking the region $\mathcal{O}_{1}$ into $\mathcal{O}_{2}$. A state $\phi$ on $\mathcal{M}\bigvee\mathcal{N}$ is separable if it is in the norm closure of the convex hull of the normal product states across $(\mathcal{M},\mathcal{N})$ i.e. it is a mixture of normal product states. If this is not so, we call $\phi$ an entangled state across $(\mathcal{M},\mathcal{N})$. Again, in therms of quantum field theoretical qubits this corresponds to the states $\phi(M)$ and $\phi(N)$ belonging to non-trivial (co)homology groups i.e. the full information about the state cannot be encoded only by means of the two subsystems separately and therefore one has to consider Mayer-Vietoris type sequences in order to restore the complete quantum state. Otherwise stated, entanglement appears as an obstruction to the merging of the actions of $M$ and $N$ via the same quantum state. In terms of worldlines, we do not have a smooth deformation taking the region $\mathcal{O}_{1}$ into $\mathcal{O}_{2}$. Only if both algebras are non-commutative i.e. quantum, can we have entangled states on the composite system. A consequence of the Reeh-Schlieder theorem [26] is that for any two non-empty sets of spacelike separated observables belonging each respectively to the space-time regions $\mathcal{O}_{1}$ and $\mathcal{O}_{2}$ with non-empty causal complements, independent on the distance between them, there exist several projections $P_{i}\in \mathcal{R}(\mathcal{O}_{i})$ which are positively correlated in the vacuum state such that $\phi(P_{1}P_{2})>\phi(P_{1})\phi(P_{2})$. This shows that the vacuum is not a product state across $(\mathcal{R}(\mathcal{O}_{1}),\mathcal{R}(\mathcal{O}_{2}))$. In order to determine if it is entangled we need a different measure called the maximal Bell correlation, defined for the pair $(\mathcal{M},\mathcal{N})$ in the state $\phi$, as 
\begin{equation}
\beta(\phi,\mathcal{M},\mathcal{N})=sup \frac{1}{2}(M_{1}(N_{1}+N_{2})+M_{2}(N_{1}-N_{2}))
\end{equation}
where the supremum is taken over all self adjoint operators $M_{i}\in\mathcal{M}$ and $N_{j}\in\mathcal{N}$ with norm less or equal to one. 
Bell inequality in the case of algebraic quantum field theory can be formulated as 
\begin{equation}
\beta(\phi,\mathcal{M},\mathcal{N})\leq 1
\end{equation}
If $\phi$ is separable across $(\mathcal{M},\mathcal{N})$ then $\beta(\phi,\mathcal{M},\mathcal{N})= 1$. It has been shown in [18], [26] and [27] that under general physical assumptions, in a vacuum representation of a local net of observables, $\beta(\phi,\mathcal{R}(\mathcal{O}_{1}),\mathcal{R}(\mathcal{O}_{2}))=\sqrt{2}$ which violates the Bell inequality. Moreover, for any spacelike separated double cones whose closures intersect i.e. tangent double cones, $\beta(\phi,\mathcal{R}(\mathcal{O}_{1}),\mathcal{R}(\mathcal{O}_{2}))=\sqrt{2}$.

\subsection{Entanglement and the Mayer-Vietoris constructions (ER $\Rightarrow$ EPR)}

\par Now that a construction capable of measuring entanglement has been designed and the observables and quantum states have been assigned each to their own space-time regions, it remains to be shown that it is possible to define entanglement as being generated by the maps of the Mayer-Vietoris sequence for a torus. 

Indeed, I showed in this article that qubits can be associated to worldlines in quantum field theory and that one- or two-qubit states can be classified in terms of (co)homology groups. Such groups will be represented by means of the basis $\{\ket{a},\ket{b}\}$. The (co)homology would then be defined by the linear combinations of elements in this basis each such combination satisfying the topological properties defining their respective (co)homology. 
The coefficients of such a linear combination belong to the coefficient structure of the cohomology. Therefore in order to work in the context of quantum mechanics the homology with complex coefficients $H_{n}(X;\mathbb{C})$ will be constructed by means of vectors $\ket{\Psi}=c_{1}\ket{a}+c_{2}\ket{b}$ with $c_{1},c_{2}\in\mathbb{C}$. This is a more suitable representation for qubits. Two-qubit states will also be classified by means of (co)homology groups but this classification may not be trivial i.e. two independent states belonging to trivial (co)homology groups may become two-qubit states belonging to non-trivial (co)homologies. This would appear as a result of the application of an entangler gate e.g. Hadamard gate on one branch followed by a two-qubit c-NOT gate. 

Now, by looking at the Mayer-Vietoris sequence one notices the appearance of direct sums of homology groups like 
\begin{equation}
H_{n}(A; \mathbb{C})\oplus H_{n}(B; \mathbb{C})
\end{equation}
Whenever the objects involved in such direct sums appear in finite numbers and represent abelian structures (like the complex numbers), the direct sums are isomorphic to the direct products and hence 
\begin{equation}
H_{n}(A; \mathbb{C})\oplus H_{n}(B; \mathbb{C})\cong H_{n}(A; \mathbb{C})\times H_{n}(B; \mathbb{C})
\end{equation}
As a basic example one may consider $\mathbb{R}\times \mathbb{R}\cong \mathbb{R}\oplus\mathbb{R}$ which both represent the cartesian plane. I will continue to use however the $\oplus$ notation for the sake of generality as, for example in the case of infinite direct sums or in the case of topological spaces with no additional structures, such an isomorphism will not apply. For all the considerations relevant to the present discussion however one may assume that $\oplus\cong \times$. 


\par What remains to be seen in what follows is that patching together a torus by means of the Mayer-Vietoris sequence implies the appearance of entanglement. To see this, one has to understand the basics of the Mayer-Vietoris method. Its original use was to detect the (co)homology of an unknown topological space $X$ by means of known (co)homologies of subspaces of $X$ which were wisely chosen such that by patching them together, the full space $X$ could be obtained. The maps capable of doing this patching formed a long exact sequence called the Mayer-vietoris sequence.
 In this article this procedure is somehow reversed, as now we know the full space is a $T_{2}$ torus and its homology is also known. We consider the two patches $A$ and $B$ on the left and the right side of the torus and form the Mayer-Vietoris sequence paying attention at the particular forms the respective maps can take. The two patches will intersect (by convention) in the upper and lower regions of the torus. The qubits belong respectively to the homologies of the patches $A$ and $B$ and, after connecting $A$ and $B$ and including them into the torus they will represent entangled qubits on the torus. 
 
It is important to notice that in the Mayer-Vietoris theorem the two groups $H_{n}(A\cap B;\mathbb{C})$ and $H_{n}(A;\mathbb{C})\oplus H_{n}(B;\mathbb{C})$ are isomorphic as groups but the inclusion maps between them do obviously not induce isomorphisms. If we look again at the Mayer-Vietoris sequence, mainly at the map $H_{n}(A\cap B;\mathbb{C})\xrightarrow{(i_{*},j_{*})}H_{n}(A;\mathbb{C})\oplus H_{n}(B;\mathbb{C})$ we notice that the map $(i_{*},j_{*})$ is induced in homology by the inclusions $i:A\cap B\hookrightarrow A$ and $j:A\cap B\hookrightarrow B$ and is not an isomorphism neither when acting on the space, nor in its homology induced form.


This map is in fact fundamental to the understanding of the dependence of entanglement on the topology, therefore we need to have it expressed in more comfortable terms. Consider therefore the standard two dimensional torus $T_{2}$ and let's start computing its second homology group by means of the Mayer-Vietoris sequence. On this path I will make the connections to entanglement as manifest as possible. For $n=2$ we have the Mayer-Vietoris sequence in the form 
\begin{widetext}
\begin{equation}
...\rightarrow H_{2}(A;\mathbb{C})\oplus H_{2}(B;\mathbb{C})\rightarrow H_{2}(T_{2};\mathbb{C})\xrightarrow{\partial}H_{1}(A\cap B;\mathbb{C})\xrightarrow{(i_{*},j_{*})}H_{1}(A;\mathbb{C})\oplus H_{1}(B;\mathbb{C})\rightarrow ... 
\end{equation}
\end{widetext}
In this part of the long sequence we can calculate all groups except the one of the torus (which however we assume it is known or at least it is not our concern to calculate it). We therefore may already write down the known parts

\begin{equation}
...\rightarrow 0 \rightarrow H_{2}(T_{2};\mathbb{C})\xrightarrow{\partial}\mathbb{C}\oplus\mathbb{C}\xrightarrow{(i_{*},j_{*})}\mathbb{C}\oplus\mathbb{C}\rightarrow ...
\end{equation}

Notice that here too, the map $(i_{*},j_{*})$ is not an isomorphism. 
 Take therefore $1$-cycles generating the homologies of $A$, $B$ and $A\cap B$ respectively in this way: for each cylinder formed by the intersection $A\cap B$ chose your cycle as the equatorial circumference. Let the associated homology classes be $\alpha$ and $\beta$. These cycles will each generate $\mathbb{C}$ and we will have 
\begin{equation}
(i_{*},j_{*}):\mathbb{C}[\alpha]\oplus \mathbb{C}[\beta]\hookrightarrow \mathbb{C}[\alpha]\oplus \mathbb{C}[\beta]
\end{equation}
but $\alpha=\beta$ when we are in $H_{n}(A;\mathbb{C})$ and $H_{n}(B;\mathbb{C})$ therefore 
\begin{equation}
(i_{*},j_{*})(\alpha,0)=(i_{*},j_{*})(0,\beta)=(\alpha,\beta)
\end{equation}
Applying a global twist in the torus (i.e. keeping the upper intersection circle unchanged and rotating the lower intersection circle around an axis perpendicular to its center by $\pi$) will not affect the physical situation but will generate the map $(i_{*},j_{*})$ which can then be written (considering the normalization factor imposed by hand in advance) as the matrix 
\[ \frac{1}{\sqrt{2}}\left( \begin{array}{cc}
1 & 1  \\
1 & -1  \\
 \end{array} \right) :\mathbb{C}\oplus\mathbb{C}\rightarrow \mathbb{C}\oplus\mathbb{C}\] 
 This matrix resulted solely from the Mayer-Vietoris theorem, a twist in the torus and a specific choice of basis but, in terms of quantum entanglement it is a standard Hadamard matrix. This matrix is used to map the qubit $\ket{0}$ into the superposition of two states with equal weight i.e. $\frac{1}{\sqrt{2}}(\ket{0}+\ket{1})$.
In terms of quantum field theoretical qubits this encodes the representation of a worldline qubit in the form of a cycle qubit.
In order to better show the analogy with quantum mechanics I detail the maps arising in the Mayer-Vietoris sequence and connect them to the hadamard-CNOT entangler gate for a bipartite system. In particular I show how the Hadamard map created by the $(i_{*},j_{*})$ inclusions is combined with the other maps arising from the Mayer-Vietoris sequence in order to produce entangled states on the two branches of a torus.
The general situation is as follows. Take two qubits $\ket{\Psi_{1}}$ and $\ket{\Psi_{2}}$ each defined in terms of quantum field theory on curved space-time as specific worldlines. In the quantum information approximation they can be seen as unit vectors each in $\mathbb{C}\times\mathbb{C}$.  For the beginning, the two states will encode both the $\ket{0}$ state. Start now with an ER space-time configuration (torus). Take the subspaces of the torus covering each one of the two handles on the left and on the right side of the torus. The intersections between these two covers occur by convention on opposing regions of the torus, let me call them the upper and the lower intersection. Let me also call the left region of the torus by $A$ and the right region by $B$.  Starting from the intersections of the two covers, the two qubits are being mapped respectively onto the two handles of the torus by means of the inclusion maps $H_{1}(A\cap B;\mathbb{C})\hookrightarrow H_{1}(A;\mathbb{C})$ and respectively $H_{1}(A\cap B;\mathbb{C})\hookrightarrow H_{1}(B;\mathbb{C})$. The upper intersection will be mapped on the left and on the right side by the map $(i_{*},j_{*})$ producing a rotated state on the upper half of the torus as if acted upon by the Hadamard gate (the normalization is introduced by hand according to the principles of quantum mechanics). The result will be $\ket{\Psi_{1}}=\frac{1}{\sqrt{2}}(\ket{0_{A}}+\ket{1_{B}})$. In general, on the lower side of the torus one can obtain similarly $\ket{\Psi_{2}}=\frac{1}{\sqrt{2}}(\ket{0_{A}}-\ket{1_{B}})$. However, to obtain the Hadamard gate (the minus sign in the last entry of the matrix) on the upper side, we used a twisted torus. This amounts basically to a change of basis. This twist will untwist the action of $(i_{*},j_{*})$ on the lower half of the torus (which would otherwise by itself try again to twist the torus) and therefore the final state on the lower torus will remain $\ket{\Psi_{2}}=\ket{0}$. 
This untwisting operation on the lower half leads to a lower map of the form 

\[ (i_{*},j_{*})= \left( \begin{array}{cc}
1 & 0  \\
0 & 0  \\
 \end{array} \right) 
 \] 
which acting on the state $\ket{0}$ leaves it unchanged (considering the convention of having $\ket{0}$ in the form of a column vector with the upper entry $1$). 

Therefore at this moment, after applying the first map of the Mayer-Vietoris sequence we obtained two qubits on the upper and lower halves of the torus
\begin{equation}
\begin{array}{cc}
\frac{1}{\sqrt{2}}(\ket{0_{A}}+\ket{1_{B}}), & \ket{0}\\
\end{array}
\end{equation}
In order to obtain the torus, the direct sum of the two homologies must be mapped in the total homology of the space. This map acts on the upper and lower components i.e. it acts on the two qubits above. This means it must be a two-qubit gate. The map is $H_{1}(A)\oplus H_{1}(B)\xrightarrow {(k_{*}-l_{*})} H_{1}(T_{2})$. The notation $(k_{*}-l_{*})$ is formal. It can be interpreted as a formal difference for the cycles of the torus but when acting on qubits it will act as a CNOT gate, as will be seen soon. The patches  have to be continuously embedded into the whole torus. But the lower side adds an extra twist via the map $(k_{*}-l_{*})$ which compensates the twist on the upper intersection (the upper intersection is not twisted by this map but it was twisted by the previous one). Therefore this map flips the second (lower) qubit when the initial first qubit has been flipped by the previous map (generating the superposition). But as the initial state was $\ket{0}$ it will only flip the lower qubit when the upper state is $\ket{1}$. Moreover, it brings us the actual homology of the torus back. Therefore what we obtained is a CNOT gate acting on two qubits, namely

\[ (k_{*}-l_{*})= \left( \begin{array}{cccc}
1 & 0 & 0 & 0 \\
0 & 1 & 0 & 0\\
0 & 0 & 0 & 1\\
0 & 0 & 1 & 0\\
 \end{array} \right) 
 \] 
(obviously, when acting on an actual qubit the proper normalization constants will be added)
Together with the previously introduced Hadamard gate the resulting state is now 
\begin{equation}
\ket{\Psi}=\frac{1}{\sqrt{2}}(\ket{0}\ket{0}+\ket{1}\ket{1})
\end{equation}
which is defined over the whole torus and therefore I can drop the indices $A$ and $B$. 
Summarizing, the quantum states after the action of the first Mayer-Vietoris map $(i_{*},j_{*})$ for the torus, are
$$
\left. \begin{array}{l}
\ket{0}\\
\ket{0}\\
\end{array}
\right\} 
\xrightarrow{(i_{*},j_{*})} 
\left\{
    \begin{array}{l}
        \frac{1}{\sqrt{2}}(\ket{0_{A}}+\ket{1_{B}})\\
        \ket{0}\\
    \end{array}
\right.
$$
As has been seen before in order to obtain an entangled state we also need the CNOT map. This map has two roles: first it has to include a second qubit in the superposed states above, second it has to switch the state of the second qubit when the first qubit is in the state $\ket{1}$ such that a truly entangled state of the two qubits emerges and third, it has to restore the whole torus from the two patches $A$ and $B$. I have shown above that such a map arises naturally from the Mayer-Vietoris sequence for a torus. For a better understanding one may have a careful look at the Mayer-Vietoris sequence
\begin{widetext}
\begin{equation}
...\rightarrow H_{1}(A\cap B;\mathbb{C})\xrightarrow{(i_{*},j_{*})} H_{1}(A;\mathbb{C})\oplus H_{1}(B;\mathbb{C})\xrightarrow{(k_{*}-l_{*})}H_{1}(T_{2};\mathbb{C})\xrightarrow{\partial} H_{0}(A\cap B;\mathbb{C})\rightarrow... 
\end{equation}
\end{widetext}
We are now interested in the map, $(k_{*}-l_{*})$. This one takes as input the sates on the two sheets covering the two handles of the torus and maps them together into a formal difference, generating the homology of the torus i.e. the vector space where the resulting entangled states will reside. While the map $(i_{*},j_{*})$ was injective, this map is surjective in order to preserve the exactness of the sequence. Merging together elements of the two sheets such that they connect in a continuous way obviously takes two qubits as an input and performs an operation on one, depending on the state of the other. These are all properties desirable for maps in the category of the CNOT map of quantum computing. The final construction I am deriving from the Mayer-Vietoris sequence is shown in fig. 1. 
\begin{figure}
\centering
\includegraphics[scale=0.7,bb=140 20 60 60]{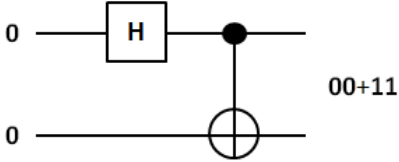}
\caption{The standard Hadamard entangler gate}
\label{fig:top}
\end{figure}
Notice first that the maps $k$ and $l$ basically map the regions $A$ and $B$ into the whole of $X=T_{2}$ after the map $(i_{*},j_{*})$ has been applied. They take the superposed state obtained after the application of the Hadamard-type map (normalization is assumed) and map it into the torus as a whole. Two aspects are important. First this will bring together the new superposition state and the original state $\ket{0}$. This basically implies tensoring the superposed qubit in the upper half with the original qubit in the lower half. Second, the two sheets must generate a torus and therefore the combination between the two maps $k$ and $l$ must be taken such that this will be the case. Formally we have 

\begin{widetext}
$$
\left. 
\begin{array}{l}
\frac{1}{\sqrt{2}}(\ket{0_{A}}+\ket{1_{B}})\\
\ket{0}\\
\end{array}
\right\} 
\xrightarrow{(k_{*},l_{*})}
        \frac{1}{\sqrt{2}}(\ket{0_{A}}+\ket{1_{B}})\otimes\ket{0}\\
\xrightarrow{(k_{*}-l_{*})} 
\begin{array}{l}
        \frac{1}{\sqrt{2}}(\ket{0}\ket{0}+\ket{1}\ket{1})\\
    \end{array}
$$
\end{widetext}


\subsection{homology with twisted coefficients, $EPR\Rightarrow ER$}
The $ER\Rightarrow EPR$ part of the duality has been derived by analyzing the form and the actions of the maps in the Mayer-Vietoris sequence of a torus. In order to make the reciprocal affirmation $EPR\Rightarrow ER$ plausible we have to explain how the entanglement of disconnected spaces (and the states defined on them) may result in a connected space. 
In general it is verified that spaces of different topology exist in mutually orthogonal sectors of the associated Hilbert space and therefore the paradox is particularly stringent. 
The connectivity of a space is determined by means of the (co)homology which, in the case of complex coefficients also represents the qubit states. 
However, when we alter the algebraic structure of the coefficients in cohomology, the information about the connectivity of a space may appear to change. Could therefore a specific non-trivial choice of coefficients lead to a non-trivial superposition of disconnected topological spaces that may result in connected topological spaces? 
We will start with two circular spaces $S^{1}$ and show that by means of a particular change in coefficients the two circular spaces representing together a disconnected space, will become a space homeomorphic to a single circle and hence a connected (although not simply connected) space. Then the resulting not simply connected space will be mapped by means of another change in coefficients into a simply connected space homeomorphic to a single point (see Fig. 2). 
The particular choice of coefficients must contain a certain twisted cyclicality. In this subsection I will discuss the process in terms of integer and twisted cyclical integer coefficients. In the next subsection a short discussion of the acyclicity of the circle will imply the use of complex coefficients [37].

\begin{figure}
\centering
\includegraphics[scale=0.25,bb=560 90 60 60]{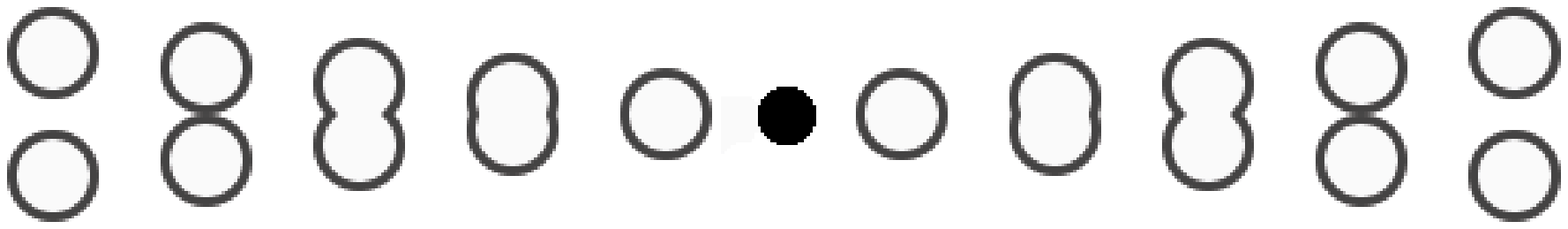}
\caption{Two circles merging, as seen by using various torsions in the coefficient groups of (co)homology. The change in the coefficient structure brings us from two independent circles to the wedge sum between two circles tangent at a common point, then to a single circle and finally to a simple point. The information is presented as seen by homology with various coefficients}
\label{fig:top}
\end{figure}

\par In order to begin, consider a circle space $S^{1}$ and an abelian group $A$. Let then $\rho:\pi_{1}S^{1}\rightarrow Aut(A)$ a representation of the fundamental group of the circle into the abelian group $A$. 

Then, the homology of the circle with coefficients in the group $A$ twisted by the map $\rho$ is $H_{k}(S^{1},A_{\rho})$. As a simple example one can consider the group $A=\mathbb{Z}_{3}$ and the the map $\rho:\mathbb{Z}\rightarrow Aut(\mathbb{Z}_{3})$ as being 
\begin{equation}
\rho = \left\{
    \begin{array}{ll}
        0\rightarrow 0 \\
        1\rightarrow 2 \\
        2\rightarrow 1 \\
        3\rightarrow 0 \\
        4\rightarrow 2\\ 
        ... 
        \end{array}
\right.
\end{equation}
The cellular chain complex associated to the homological representation of the circle is then 
\begin{equation}
0\rightarrow \mathbb{Z}[t,t^{-1}]\xrightarrow{\delta}\mathbb{Z}[t,t^{-1}]\rightarrow 0
\end{equation}
$\delta$ is the boundary map which by definition represents the multiplication with $(t-1)$. Therefore $t$ and $t^{-1}$ define the required ring structure for the circular space. We therefore have an isomorphism $\mathbb{Z}[\pi_{1}S^{1}]\cong\mathbb{Z}[t,t^{-1}]\cong\mathbb{Z}[\mathbb{Z}]$ which will slightly simplify the calculation without affecting the final result. Let me now tensor with $\mathbb{Z}_{3}$ in order to obtain the homology with the desired coefficients over $\mathbb{Z}[t,t^{-1}]$. Then I obtain 

\begin{widetext}
\begin{equation}
\mathbb{Z}_{3}\xrightarrow{\cong}\mathbb{Z}[t,t^{-1}]\otimes_{\mathbb{Z}[t,t^{-1}]}\mathbb{Z}_{3}\xrightarrow{\delta \otimes Id}\mathbb{Z}[t,t^{-1}]\otimes_{\mathbb{Z}[t,t^{-1}]}\mathbb{Z}_{3}\xrightarrow{\cong}\mathbb{Z}_{3}
\end{equation}
\end{widetext}

The first map is $a \rightarrow 1\otimes a$ and the last map is $1\otimes a\rightarrow a$. It is required to reduce to $1\otimes a$ before applying the last map. The result therefore is

\begin{equation}
a\rightarrow 1\otimes a\rightarrow (t-1)\otimes a=1\otimes (ta-a)\rightarrow t a-a
\end{equation}

The boundary map obtained after tensoring with $\mathbb{Z}_{3}$ is then 
\begin{equation}
D:\mathbb{Z}_{3}\rightarrow\mathbb{Z}_{3}
\end{equation}
\begin{equation}
\begin{array}{l}
D(0)=0\\
D(1)=t\cdot 1-1=2-1=1\\
D(2)=t\cdot 2 - 2 = 1-2=2\\
\end{array}
\end{equation}
and hence is the identity on $\mathbb{Z}_{3}$. 
Therefore the homology groups of $S^{1}$ with coefficients in $\mathbb{Z}_{3}$ twisted by the nontrivial map $\rho$ are all trivial
\begin{equation}
H_{0}(S^{1};\mathbb{Z}_{3})_{\rho}\cong H_{1}(S^{1};\mathbb{Z}_{3})_{\rho}\cong...\cong 0
\end{equation}
This shows how a circle can be mapped into a point via a controllable change of coefficients in homology provided all information obtained about the space is obtained via (co)homology. 
Let me further apply a similar procedure that will merge two disjoint circles into one single circle. 
In order to do this the coefficient group $A$ will now be $\mathbb{Z}_{2}$ and the twisting will have the form 
$$
\rho = \left\{
    \begin{array}{ll}
        0\rightarrow 1 \\
	    1\rightarrow 0 \\
	    2\rightarrow 1 \\
	    3\rightarrow 0 \\
        ... \\
     \end{array}
\right.
$$
The analyzed space will now be a disjoint union of circles $S^{1}$ namely $X=S^{1}\sqcup S^{1}$. By a simple application of Mayer-Vietoris theorem it results that $H_{q}(S^{1})\cong H_{q}(S^{1})\oplus H_{q}(S^{1})$. Now, by using the twisted coefficients as described above, the homology won't be able to distinguish the two circles and hence we arrive at the single circle case.

\subsection{More twisted coefficients, $EPR \Rightarrow ER$}

\par It appears that the ``quantum superposition" of topological spaces may be governed by a deeper form of entanglement, one in which the role of the linear superposition is altered by the structure of the coefficient ring in (co)homology. While keeping the formal linear combinations of subspaces or states as defined in normal quantum mechanics, changing the algebraic structure of the coefficients of such combinations (a prescription that amounts to the change of the algebraic structure of the coefficients in cohomology and implicitly to the addition of topological structure to the previously trivial mathematical point) allows us to explore global statistical phenomena that make entanglement visible. However, it is clear that there is no linear quantum observable that can be associated to entanglement and therefore entanglement itself is not a linear phenomenon [4], [56]. Therefore, by employing different coefficient structures one may entangle topologically disconnected pieces of space-time producing (not necessarily simply) connected space-times if certain restrictions on the coefficient structures are being imposed. The obvious result is that the topology of spacetime is an emerging feature guided by entanglement. This new form of entanglement (resulting from linear combinations with coefficients of non-trivial algebraic structures) is governed by the universal coefficient theorem in the sense that it allows us to switch from the information which can be obtained by means of one coefficient structure to the information obtainable via the other coefficient structure. Like in the case of normal entanglement, some questions about the topological space cannot be meaningfully answered when one relies exclusively on one coefficient structure. Therefore, entanglement as a linear combination of topological spaces in this case admits extra-flexibility due to the various possible choices of coefficient rings and the global effects such choices entail. This cannot be ignored because in this case the coefficient rings may alter the topological information which can be extracted from the given spaces. 
Therefore in this final chapter I briefly extend the analogy between qubits and homological algebra by going to a (co)homology theory with twisted complex coefficients. The key property of twisted (co)homology is the twisted acyclicity of the circle [37]. This property tells us that a twisted homology of a circle with coefficients in $\mathbb{C}$ which have a non-trivial monodromy must vanish. Subsequently a twisted homology theory of this kind completely ignores the parts of the space it wishes to describe which are formed by circles along which the monodromy of the coefficient system is non-trivial. The implications to physics are important mainly because, as I argued in [17], the use of coefficient systems of various forms and of the universal coefficient theorem amounts to a prescription of finding new dualities in physics i.e. different analytical tools used to describe the same phenomena. In this case the duality is between entanglement and topology. In general for a homology theory, the dimension of $H_{0}(X;\mathbb{C})$ is equal to the number of path-connected components in $X$. Also, in classical homology theory (based on the standard Eilenberg-Steenrod axioms) $H_{0}(X;\mathbb{C})$ does not vanish unless $X$ is empty. For twisted homology this last property is not valid anymore. Particularly when we analyze a circle $X=S^{1}$, we consider the map $\mu: H_{1}(S^{1})\rightarrow C^{\times}$ taking the generator $1\in\mathbb{Z}=H_{1}(S^{1})$ to $\zeta\in \mathbb{C}^{\times}$. By this twist we then have the acyclicity of the circle in the sense that $H_{*}(S^{1};\mathbb{C}^{\mu})=0$ if and only if $\zeta\neq 1$. Moreover, let $X$ be a path connected space and $\mu:H_{1}(S^{1}\times X)\rightarrow \mathbb{C}^{\times}$ be a homomorphism. Then let $\zeta$ be the image under $\mu$ of the homology class realized by a fiber $S^{1}\times pt$. Then $H_{*}(S^{1}\times X;\mathbb{C}^{\mu})=0$ if $\zeta\neq 0$. The proof of these results can be found in [37]. Physically this means that we may consider quantum states on a region of our space as belonging to the homology with complex coefficients $\ket{\Psi}\in H_{1}(X;\mathbb{C})$. $X$ is in this case is the direct sum of two disconnected regions $X=A \sqcup B$. The homology of such a space will be the direct sum of the homologies of the two disjoint regions $H_{1}(X;\mathbb{C})= H_{1}(A;\mathbb{C})\oplus  H_{1}(B;\mathbb{C})$.  We can choose $A$ and $B$ to be spacelike separated. The state $\ket{\Psi}$ is entangled over $A$ and $B$ although the space itself doesn't show any topological features at this moment. The same properties will remain valid when we change the coefficient structure $\mathbb{C}\rightarrow\mathbb{C}^{\mu}$ where the twisting induced by $\mu$ is such that the coefficients form a twisted system with a non-trivial monodromy around any circle connecting region $A$ and $B$. But with such coefficients $H_{1}(X;\mathbb{C})$ becomes trivial and hence the two regions become trivially identified i.e. in a sense similar to quantum teleportation. However, we can now modify the space $X$, by introducing the required circles which will make it look like a torus. This cannot affect the homology with twisted coefficients as it is not sensitive to circular components. However, if we now move back to untwisted coefficients we need to carefully employ the universal coefficient theorem and we will obtain the standard homology of a torus in complex coefficients, particularly $H_{1}(X;\mathbb{C})=\mathbb{C}\oplus\mathbb{C}$. Summarizing, we started with a flat space and an entangled state and by changing the coefficient structure to a twisted one, making some undetectable changes to the space which left the homology intact and then changing back to the original coefficients we obtained the homology of a torus in complex coefficients. Of course the last transformation cannot be performed without penalizing some bijective maps due to the universal coefficient theorem. However, the physically relevant states remain unchanged, the only modifications being at the level of the $Tor$ and $Ext$ functors arising in the universal coefficient theorem for homology respectively cohomology. But how can it be that the physical states obtained when we go back to complex coefficients do not match the original states (as we do not have an absolute bijection because of the $Tor$ and $Ext$ functors)? First one should notice that $Ext$ and $Tor$ encode precisely the deviations introduced by adding the circular components. Therefore, this collapse of the bijection is simply because to begin with we made an assumption which cannot hold after a proper topological analysis, namely that in the original case we have a flat, topologically trivial space-time and entangled states. The whole point of this article is to show that such a situation is impossible, as entanglement automatically has to imply non-trivial space-time topologies. The main result is that entanglement is precisely encoded in the homology of a torus and a torus precisely encodes entanglement but entanglement cannot exist in topologically trivial space-time. It is obviously interesting to interpret this result in the case of basic quantum entanglement experiments where, apparently, the topology of space-time changes. How should such a change be interpreted in terms of basic entanglement experiments and apparently flat space-time remains a mystery, although mathematically it is possible to have a flat, topologically non-trivial space-time.


\section{7. Reeh-Schlieder theorem and the ER-EPR duality}
The example of the previous section indicates that the Mayer-Vietoris theorem plays an important role in the characterization of the ER-EPR duality. However, several transformations done there may appear somewhat artificial. In order to strengthen the argument in favor of the ER-EPR duality a more general approach is needed. Therefore, now I will relate the maps arising in the Mayer-Vietoris sequence with the Reeh-Schlieder theorem. Before entering a more detailed analysis, let me briefly summarize the present strategy. 

The Reeh-Schlieder theorem can be seen as a generalized statement of the state-operator correspondence from conformal field theory. This correspondence basically states that there exists a bijective relation between the operators of the theory strictly localized at one point and all the quantum states of the theory. Such a correspondence is somehow counterintuitive as this would mean that all the local operators are to be put into a bijective correspondence to states defined basically over the entire space under consideration. This result depends on the existence of conformal symmetry. However, for a generic quantum field theory a similar theorem exists, albeit the map is now surjective, i.e. one may always map operators to states but not every state corresponds uniquely to a single local operator. The result for general quantum field theories however is also important as it states that any quantum state on the considered space can be generated by applying a local operator on the vacuum. This means that a quantum state, spatially separated from the region where the local operator is defined can also be created by the action of that same local operator. This is generally interpreted as a quantum field theoretical manifestation of entanglement and is basically the general formulation of the Reeh-Schlieder theorem in quantum field theory. In order to connect this to the Mayer-Vietoris theorem one must focus on the $(k_{*}-l_{*})$ map defined in the previous chapter. In this context one defines the localized operators as belonging to the regions $A$ and $B$. The operators themselves are classified by the homology groups of the two regions albeit individually they are all strictly localized. I show that the map $(k_{*}-l_{*})$ defined above induces basically the same result as the Reeh-Schlieder theorem i.e. quantum field theoretical entanglement. In order for this analogy to be plausible it is important to consider the states as being defined over the whole space (in this case the torus) and as being classified by the resulting total homology group. Therefore, the $(k_{*}-l_{*})$ map relates the operators localized in the regions $A$ and $B$ to the states defined over the entire space or in regions spatially separated from where the operators are defined. From the exactness of the Mayer-Vietoris sequence one can notice that in general this map is a surjection i.e. every local operator may be mapped into a state but more operators may correspond to the same state. For this map to become a bijection the total homology of the space should become trivial. But the trivialization of the homology groups will be equivalent to reducing the torus to a single point. This is precisely what would happen if we mapped the states on the cylinders into the initial state (central point) of a radially quantized conformal field theory. This is a particularity of conformal field theories not generalizable to other quantum field theories. The state-operator correspondence in radially quantized conformal field theory states precisely this i.e. every state of the field theory can be generated by employing the operators of the theory all localized at the center. It is somehow surprising to notice that reducing the theory under consideration to a conformal field theory amounts to a choice of twisted cyclic coefficients in the homology. In that case, as I showed in the last part of the previous section, we also reduce the cyclic components of a space to points.This type of dualities will be further discussed in another article. 
\subsection{The Reeh-Schlieder theorem}
Quantum field theories are characterized by the ubiquity of fluctuations and of long-range correlations. Moreover, using suitable selective operations and applying them in a localized but arbitrary region of the spacetime vacuum, any given state can be created but not only in that particular region but in any other causally separated spacetime region. This result is known as the Reeh-Schlieder theorem [42]. To describe it in a more rigorous form consider a spacetime manifold $M$ and a family of local operators $\{\mathcal{A}(\mathcal{O})\}_{\mathcal{O}\subset M}$ forming a $C^{*}$ algebra, all acting on a Hilbert space $\mathcal{H}$. The family is considered to be indexed by the open subsets of $M$ subject to conditions of isotony and locality
\begin{equation}
\begin{array}{cc}
\mathcal{O}_{1}\subset \mathcal{O}\Rightarrow \mathcal{A}(\mathcal{O}_{1})\subset\mathcal{A}(\mathcal{O}),\; & \mathcal{O}_{1}\subset\mathcal{O}^{\perp}\Rightarrow \mathcal{A}(\mathcal{O}_{1})\subset \mathcal{A}(\mathcal{O})'
\end{array}
\end{equation}
The set of all points of $M$ which cannot be connected to $\mathcal{O}$ by any causal curve are here called $\mathcal{O}^{\perp}$. $\mathcal{A}(\mathcal{O})'$ denotes the commutant algebra of $\mathcal{A}(\mathcal{O})$ over the set of all operators acting on the Hilbert space $B(\mathcal{H})$. Now, one can say that a unit vector $\Omega\in\mathcal{H}$ satisfies the Reeh-Schlieder property with respect to the region $\mathcal{O}\subset M$ if $\Omega$ is cyclic for the algebra $\mathcal{A}(\mathcal{O})$ of observables localized in $\mathcal{O}$. This means that the set of vectors $\mathcal{A}(\mathcal{O})\Omega=\{A\Omega:A\in \mathcal{A}(\mathcal{O})\}$ is dense in $\mathcal{H}$. Otherwise stated, the local operator is sufficient to generate the whole Hilbert space. One also says that $\Omega$ has the Reeh-Schlieder property if $\Omega$ is cyclic for $\mathcal{A}(O)$ for each $\mathcal{O}\subset M$ which is open, non-void and relatively compact. If one considers the locality assumption as well, this also implies that $\Omega$ is separating for all local algebras $\mathcal{A}(\mathcal{O})$ i.e. $A\Omega=0\Rightarrow A=0$ for all $A\in\mathcal{A}(\mathcal{O})$. Generalizations of the Reeh-Schlieder theorem have been constructed for curved spacetime [43, 44, 45]. 
The Reeh-Schlieder theorem is also responsible for the violation of Bell's inequalities in quantum field theory [46] and for long range entanglement of states in relativistic quantum field theory [47, 48, 49]. The requirement that the spacetime in which the quantum system evolves has some specific isometries for the Reeh-Schlieder theorem to be valid was relaxed in the work published in [50, 51]. In what follows I will show that the results of the Reeh-Schlieder theorem, mainly those entailing long range entanglement, can be analogously described by means of the Mayer-Vietoris theorem applied to a torus. 
\subsection{Mayer-Vietoris and Reeh-Schlieder}
As noted in the previous subsection, the Reeh-Schlieder theorem is applicable in the most general situations and entails vacuum correlations and quantum entanglement. Can the Mayer-Vietoris theorem be employed to arrive at results analogous to those of Reeh-Schlieder? Apparently the answer to this question is yes and I will argue for that in what follows. The main feature of Reeh-Schlieder is that every quantum state can be constructed by means of local operators, even those quantum states spatially separated from the regions where the local operators reside. To see how this happens let us go back to the Mayer-Vietoris sequence for the torus and consider now the localized operators as being defined on the two sheets $A$ and $B$ such that 
\begin{widetext}
\begin{equation}
\begin{array}{cccc}
A\subset M,\; & B\subset M,\; & \mathcal{A}(A)\Omega_{A}=\{Q_A\Omega_{A} : Q_A\in\mathcal{A}(A)\},\; & \mathcal{A}(B)\Omega_{B}=\{Q_B\Omega_{B} : Q_B\in\mathcal{A}(B)\}
\end{array}
\end{equation}
\end{widetext}
The algebras of operators can also be classified by means of the homology groups of the spaces (or sheets) where they are localized. Therefore we are studying again the homology groups by means of the Mayer-Vietoris sequence. The focus is now on the map $(k_{*}-l_{*})$. Its role is to patch together the two regions where the local operators are defined such that they form the whole torus $H_{1}(A;\mathbb{C})\oplus H_{1}(B;\mathbb{C})\xrightarrow{(k_{*}-l_{*})} H_{1}(T_{2};\mathbb{C})$. Obviously the quantum states can also be classified by the homology groups. In particular the quantum states defined over the entire torus can be classified by the homology $H_{1}(T_{2};\mathbb{C})$ but not completely by each of the homologies $H_{1}(A;\mathbb{C})$ or $H_{1}(B;\mathbb{C})$. 
Therefore, while operators in $H_{1}(A;\mathbb{C})$ may act on states classified by $H_{1}(A;\mathbb{C})$ they must be able to produce states defined also outside the domain of states classified by $H_{1}(A;\mathbb{C})$. This can be seen by the fact that the map $(k_{*}-l_{*})$ maps operators belonging to $H_{1}(A;\mathbb{C})$ into the homology which classifies the states defined over the whole torus namely $H_{1}(T_{2};\mathbb{C})$. The same is also obviously valid for $H_{1}(B;\mathbb{C})$ and for the direct sum $H_{1}(A;\mathbb{C})\oplus H_{1}(B;\mathbb{C})$. Moreover, due to the exactness of the Mayer-Vietoris sequence this map will have to be surjective. Therefore no quantum state classified by $H_{1}(T_{2};\mathbb{C})$ will remain uncovered by an operator from region $A$, from region $B$ or from their direct sum. One could argue now that this would not mean all global states are generated by means of localized operators only, as the direct sum practically involves all the operators, on both sides $A$ and $B$. This would not be correct. To see why, one should look at the previous arrow in the sequence, namely $H_{1}(A\cap B;\mathbb{C})\xrightarrow{(i_{*},j_{*})} H_{1}(A;\mathbb{C})\oplus H_{1}(B;\mathbb{C})$. This arrow is injective and maps operators in the intersection of the two regions $A\cap B$ into the direct sum.
By the convention of the previous chapter, the intersection $A\cap B$ represents two disjoint region on the upper and lower sides of the torus. The inclusion maps which bring the operators from this region into the direct sum of the homologies of $A$ and $B$ takes every element in the homology of the intersection. However, it doesn't cover all the elements of $A$ and $B$. But in this case the homology classes are important and the elements in them can be homologically transformed one into the other. 
Therefore, at the level of homologies, each operator from $H_{1}(A\cap B;\mathbb{C})$ can be mapped into operators of $H_{1}(A;\mathbb{C})\oplus H_{1}(B;\mathbb{C})$. At the origin of this sequence however, all operators were basically localized in the upper and lower regions of the torus, namely $(A\cap B)$. We were therefore able to recover all states that can be defined on the torus i.e. $H_{1}(T_{2};\mathbb{C})$ using only the upper and lower domains defined by $A\cap B$ and a set of maps defined by the Mayer-Vietoris sequence.  But such a relation between strictly localized operators and states defined over the whole space is precisely the result of the Reeh-Schlieder theorem i.e. localized operators acting on the vacuum are capable of generating every quantum state associated to the theory over the whole space. 
Therefore, the Mayer-Vietoris theorem has been related to the Reeh-Schlieder theorem which basically encodes entanglement as described in quantum field theory. Another property of quantum field theories derivable from the Reeh-Schlieder theorem is the entanglement of the vacuum. I will not insist on this aspect here, as it has been extensively discussed in [52, 53]. However, due to this newly proved connection, topology appears to be a useful tool for the description of the entanglement of vacuum in quantum field theories. A connection to the entanglement entropy will also be the subject of a future research.

\section{8. Conclusions}
Finally, some conclusions are in order. The ER-EPR duality is a widely reaching observation based on numerous physical facts. However, it remains an unproved statement. There are various paths that may lead to the conclusion that the duality is valid. Checking the entanglement of vacuum states is certainly one such path. However, it appears that the computational details of such an endeavor may mask some more subtle and general observations. Therefore, I adopted in this paper a different path, one that starts looking at the ER-EPR phenomenon from a different perspective, mostly related to homological algebra and algebraic topology. It results that the connection between topology and entanglement is deeper that thought previously. In articles like [28], [29] the authors explore the connection between topological and quantum entanglement at various levels of accuracy and precision. Indeed, Borromean rings and braid groups as examples of topological entanglement may partially encode some aspects of quantum entanglement. However, it appears now that quantum entanglement has far deeper roots, originating in homological algebra and not being truly dependent on the linking or braiding structures themselves. 
The main result of this paper is that an ubiquitous theorem of homological algebra gives new and unexpected insights on the nature of entanglement and its relation to algebraic topology. It appears that in the process of patching together spaces of more complicated topological structure out of simpler objects, usually with well known topological invariants, entanglement is emerging and actually becomes unavoidable. In this sense basic procedures that help us understand the topology of more complicated spaces in terms of the topology of simpler spaces are at the foundation of what we know in quantum mechanics or quantum field theory as entanglement. Therefore, entanglement is not only a geometrical or even a topological property, but instead, a property that emerges from the procedures required in order to construct spaces of a higher degree of complexity. The simplest case, a sphere, to be associated with the flat space-time of basic quantum field theory may already be seen as entangled due to the combination of two sheets covering each one hemisphere. Adding further complexity only re-confirms the requirement of entanglement which appears to be rather ubiquitous in both physics and topology. 
The simplest structure presented in this article is obviously a simplification, albeit a quite suggestive one. Further complications may appear if a string-theoretical object may be considered for example, at the heart of entangled black holes. Topologies may alter in different ways and truly quantum gravitational computations will have to take them all into account via whatever is the quantum-gravity analogue of a quantum amplitude. One component of this amplitude may be a system formed by gluing two hyper-spheres at a point or by considering $AdS$ spaces and decomposing them into simpler pieces.


\begin{thebibliography}{99}
\bibitem{1} E. Schrodinger, "Discussion of Probability Relations Between Separated Systems", Proceedings of the Cambridge Philosophical Society 31 (1935), 555-662
\bibitem{2} M. V. Raamsdonk, Gen. Rel. Grav. 42, 2323-2329 (2010)
\bibitem{3} St. J. Summers, R. Werner, Phys. Lett. 110A, No. 5 (1985)
\bibitem{4} J. Maldacena, L. Susskind, Fortsch. Phys. 61: 781-811(2013)
\bibitem{5} L. H. Kauffman, S. J. Lomonaco Jr. New J. Phys. Vol. 4, 73, (2002)
\bibitem{6} D. Zhou, G. W. Chern, J. Fei, R. Joynt, Int. J. Mod. Phys. B 26, 1250054 (2012)
\bibitem{7} T. P. Oliveira, P. D. Sacramento, Phys. Rev. B 89, 094512 (2014)
\bibitem{8} A. Hamma, W. Zhang, S. Haas, D. A. Lidar, Phys. Rev. B 77, 155111 (2008)
\bibitem{9} F. S. N. Lobo, G. J. Olmo, D. Rubiera-Garcia, The European Physical Journal C, 74, 2924, (2014)
\bibitem{10} Y. Zhang, T. Grover, A. Turner, M. Oshikawa, A. Vishwanath, Phys. Rev. B 85, 235151 (2012)
\bibitem{11} L. H. Kauffman, S. J. Lomonaco Jr,  New J. Phys, Volume 6, 2004
\bibitem{12} L. Vietoris, Monatshefte fur Mathematik 37, pag. 159Ð62, (1930)
\bibitem{13} F. Mintert, C. Viviescas, A. Buchleitner, Lect. Notes Phys. 768, 61-86 (2009)
\bibitem{14} P. Zanardi, D. Lidar, S. Lloyd, Phys. Rev. Lett. 92, 060402, (2004)
\bibitem{15} L. Derkacz, M. Gwozdz, L. Jakobczyk J. Phys A, Vol. 45, No. 2
\bibitem{16} St. J. Summers, R. Werner, J. Math. Phys. 28, 2440 (1987)
\bibitem{17} A. T. Patrascu, Phys. Rev. D 90, 045018 (2014)
\bibitem{18} St. J. Summers, R. Werner, Commun. Math. Phys. 110, pag. 247-259 (1987)
\bibitem{19} M. D. Kruskal, Phys. Rev. 119, 5, pag. 1743 (1960)
\bibitem{20} G. Vidal, J. I. Latorre, E. Rico, A. Kitaev, Phys. Rev. Lett. 90, 227902 (2003)
\bibitem{21} A. Kitaev, J. Preskill, Phys. Rev. Lett. 96, 110404 (2006) 
\bibitem{22} O. Ben-Ami, D. Carmi, M. Smolkin, Journal of High Energy Physics, 2015:48 (2015)
\bibitem{23} A. Strominger, C. Vafa, Phys. Lett. B, 379, 99 (1996)
\bibitem{24} S. Ryu, T. Takayanagi, Phys. Rev. Lett. 96, 181602 (2006)
\bibitem{25} F. M. Haehl, T. Hartman, D. Marolf, H. Maxfield , M. Rangamani, Journal of High Energy Physics, 2015:23 (2015)
\bibitem{26} S. Schlieder, Comm. Math. Phys. 1, No. 4, pag. 265-280 (1965)
\bibitem{27} B. Reznik, arXiv : quant-ph/0008006
\bibitem{28} A. Sugita, Proc. Int. Workshop on "Knot Theory for Scientific Objects", Osaka, Japan, (2006)
\bibitem{29} F. L. Thorp-Greenwood, A. N. Kulak, M. J. Hardie, Nature Chemistry 7, 526-531 (2015)
\bibitem{30} A. Hatcher, Algebraic Topology, Cambridge University Press, ISBN 0-521-79540-0  (2002)
\par for the definition of simplicial complexes see pag. 106
\par for the definition of homology with non-trivial coefficients see pag. 153
\bibitem{31} M. D. Iftime, Observables and cohomology classes for classical gravitational field, arXiv:0809.3596 [gr-qc] (2008)
\bibitem{32} A. Schwarz, Lett. Math. Phys. 49, 2, pag. 115 (1999)
\bibitem{33} H. Sati, J. of Geom. Phys. 62, 5, pag. 1284 (2012)
\bibitem{34}A. Peres, D. R. Terno, Rev. Mod. Phys., 76, pag. 93 (2004)
\bibitem{35}D. R. Terno, Phys. Rev. Lett., 93, 051303 (2004)
\bibitem{36}M. B. Plenio, J. Eisert, J. Dreissig, M. Cramer, Phys. Rev. Lett. 94, 060503 (2005)
\bibitem{37}O. Viro, Journal of Knot theory and its Ramifications, 18, pag. 729 (2009)
\bibitem{38} R. Colella, A. W. Overhauser, S. A. Werner, Phys. Rev. Lett., 34, pag 1472 (1975)
\bibitem{39} S. A. Werner, Class. Quant. Grav., 11(6A), pag. 207 (1994)
\bibitem{40} J. Anandan, Phys. Rev. D, 15, pag. 1448 (1977)
\bibitem{41} J. Audretsch, C. Lammerzahl, Appl. Phys. B, 54, pag. 351 (1992)
\bibitem{42} H. Reeh, S. Schlieder, Nuovo Cimento, 22, 5, pag. 4751 (1961)
\bibitem{43} A. Strohmaier, R. Verch, M. Wollenberg, J. Math. Phys., 43, pag. 5514 (2002)
\bibitem{44} D. Buchholz, O. Dreyer, M. Florig, S. J. Summers, Rev. Math. Phys., 12, pag. 475 (2000)
\bibitem{45} D. Buchholz, J. Mund, S. J. Summers, Fields Inst. Comm., 30, pag. 65 (2001)
\bibitem{46} S. J. Summers, R. Werner, Ann. Inst. H. Poincare, 49, pag. 215 (1988)
\bibitem{47} H. Halvorson, R. Clifton, J. Math. Phys., 41, pag. 1711 (2000)
\bibitem{48} C. D. Jakel, Found. Phys. Lett., 14, pag. 1 (2001)
\bibitem{49} R. Verch, Lett. Math. Phys., 28, pag. 143 (1993)
\bibitem{50} C. D. Jakel, J. Math. Phys., 41, pag. 1745 (2000)
\bibitem{51} R. F. Streater, A. S. Wightman, "PCT, Spin and Statistics, and all that", 2nd ed., Addison-Wesley (1989)
\bibitem{52} C. D. Jakel, Ann. d. Physik, 12, 5, pag. 289 (2003)
\bibitem{53} I. Ibnouhsein, F. Costa, and A. Grinbaum, Phys. Rev. D, 90, 065032 (2014)
\bibitem{54} M. C. Palmer, M. Takahashi, H. F. Westman, Annals of Physics, 327, pag. 1078 (2012)
\bibitem{55} M. C. Palmer, M. Takahashi, H. F. Westman, Annals of Physics, 336, pag. 505 (2013)
\bibitem{56}N. Bao, J. Pollack, G. N. Remmen, Journal of High Energy Physics, JHEP 1511:126 (2015)
\bibitem{57}A. T. Patrascu, Condens. Matter 2017, 2, 13
\end{thebibliography}
\end{document}